\documentstyle[aps,preprint,prl,epsf]{revtex}
\tightenlines

\newcommand{\be}{\begin{equation}}
\newcommand{\ee}{\end{equation}}

\newcommand{\bmath}{\begin{mathletters}}
\newcommand{\emath}{\end{mathletters}}

\begin{document}

\title{\Large{\bf Correlation Effects in Band Ferromagnetism}}

\vskip0.5cm 

\author{ G. G\'{o}rski and J. Mizia }

\address{Institute of Physics, University of Rzesz\'{o}w, ulica Rejtana 16A, \\
35-958 Rzesz\'{o}w, Poland\\}

\maketitle

\vskip0.5cm 


\begin{abstract}

\noindent
We analyze the influence of different on-site and inter-site interactions on the Curie temperature of transition metal magnetic elements. 
The numerical calculations show the well known result that the on-site Coulomb repulsion helps ferromagnetism mainly at half-filling of the band, where the 3d elements are antiferromagnetic (Cr, Mn). The inter-site interactions, which in the new approximation decrease the width of the band, favor ferromagnetism at both ends of the band. At the same time they lower the ferromagnetic Curie temperature towards experimental values, removing the paradox which persisted for a long time.

\end{abstract}

\vskip0.5cm

 


\newpage

\noindent {\Large {\bf 1. Introduction } }

\vskip0.5cm

Recently there has been a resurgence of research \cite{1,2,3,4,5,6,7,8} into the influence of on-site and inter-site interactions on band magnetism. The authors of this paper have shown how different inter-site interactions would affect the ferromagnetism in the presence of the on-site Coulomb repulsion. It has been shown \cite{1,7} that the average of two electron operators on neighboring lattice site is proportional to the band's kinetic energy, it affects the kinetic energy, and it can be calculated in the simple way. The results have decreased the magnitude of interaction constants necessary to create the ferromagnetism.
 
We have also shown \cite{9} that the simple rectangular and parabolic density of state (DOS) cannot represent the
itinerant magnetic elements. The rectangular DOS does not have a unique relationship between the Stoner field and
magnetization, and the parabolic DOS does not allow for taking into account the inter-site interactions. The simplest possible DOS, which can be used, is the semi-elliptic DOS.

In this paper we will use this new many body technique (mentioned above) in combination with the simple semi-elliptic DOS, for calculation of the Curie temperature of basic ferromagnetic elements.

\vskip1.0cm 

\noindent {\Large {\bf 2. The Model}} 

\vskip0.5cm 
 
We follow closely our previous paper \cite{7} and we employ the following model Hamiltonian 

\be
 \begin{array}{c}
 H=-\sum\limits_{<ij>\sigma } {t_{ij}^\sigma \left( {c_{i\sigma }^+ c_{j\sigma 
} +h.c.} \right)}-\mu _0\sum\limits_i {\hat {n}_i } -F \sum\limits_{i\sigma } {n_\sigma \hat 
{n}_{i\sigma } } +U\sum\limits_{i\sigma } {\hat {n}_{i\sigma } \hat 
{n}_{i-\sigma } } \\ 
 +V\sum\limits_{<ij>} {\hat {n}_i \hat {n}_j } +J\sum\limits_{<ij>\sigma 
,\sigma '} {c_{i\sigma }^+ c_{j\sigma '}^+ c_{i\sigma '} c_{j\sigma } } 
+J'\sum\limits_{<ij>} {\left( {c_{i\uparrow }^+ c_{i\downarrow }^+ 
c_{j\downarrow } c_{j\uparrow } +h.c.} \right)} \\ 
 \end{array},
\label{1}
\ee
where $\mu_0$ is the chemical potential, $c_{i\sigma }^+({c_{i\sigma } })$ creates (annihilates) the electron with
spin $\sigma$ on the $i$th lattice site, $\hat n_{i\sigma }  = c_{i\sigma }^ +  c_{i\sigma }$ is the particle
number operator for electrons with spin  $\sigma$  on the $i$th lattice site, $\hat n_i  = \hat n_{i\sigma }  + 
\hat n_{i - \sigma }$ is the charge operator, $U$ is the on-site Coulomb repulsion and $F$ is the intra-atomic
Hund's interaction. In the Hamiltonian (1) we have three explicit inter-site interactions; $J$ -exchange interaction, $J'$ -pair hopping interaction, $V$ -density-density interaction. The spin dependent correlation hopping  $t_{ij}^\sigma$ depends on the occupation of sites $i$ and $j$, and in the operator form it can be expressed as  

	\be
t_{ij}^\sigma   = t (1 - \hat n_{i - \sigma } )(1 - \hat n_{j - \sigma } ) + t_1 \left[ {\hat n_{i - \sigma } (1 - 
\hat n_{j - \sigma } ) + \hat n_{j - \sigma } (1 - \hat n_{i - \sigma } )} \right] + t_2 \hat n_{i - \sigma } \hat 
n_{j - \sigma },
\label{2} 
\ee 
where $t$ is the hopping amplitude for an electron of spin  $\sigma$ when both sites $i$ and $j$ are empty.

Parameters $t_1$, $t_2$  are the hopping amplitudes for an electron of spin $\sigma$  when one or both of the sites
$i$ or $j$ are occupied by an electron with opposite spin, respectively. Including the occupationally dependent
hopping given by Eq. (2) into the Hamiltonian (1) we obtain the following result 
	
	\be
	\begin{array}{c}
 H=  -\sum\limits_{<ij>\sigma } {\left[ {t_0 -\Delta t\left( {\hat {n}_{i-\sigma 
} +\hat {n}_{j-\sigma } } \right)+2t_{ex} \hat {n}_{i-\sigma } \hat 
{n}_{j-\sigma } } \right]\left( {c_{i\sigma }^+ c_{j\sigma } +h.c.} \right)}-\mu _0 \sum\limits_i {\hat {n}_i } +\\ 
\text{+potential energy} 

 \end{array},
 \label{3}
\ee	  
	where
	\be
	\Delta t = t - t_1   ,\qquad  
t_{ex}=\frac{{t + t_2 }}{2} - t_1 .
\label{4}
  \ee  

In this form it is quite visible that the kinetic interactions:  $\Delta t$ and $t_{ex}$ are also the inter-site interactions.

We assume that $t_1/t=S$ and $t_2/t_1=S_1$. In general these parameters are different and they both fulfill the condition $S<1$  and $S_1<1$ which is equivalent to $t>t_1>t_2$ (see Ref. \cite{10}). For simplicity we will assume
now that $S\equiv S_1$ . With this assumption we have the relationship 
 \be
	\Delta t = t - t_1=t(1-S)   ,\qquad  
t_{ex}=\frac{{t + t_2 }}{2} - t_1=\frac{{t}}{2} .
\label{5}
  \ee  			

In the Hamiltonian (3) there are many inter-site interactions: $\Delta t$, $t_{ex}$, $J$, $J'$, $V$, for which we
will use the modified Hartree-Fock (H-F) approximation. The main point of this approximation is to retain the
following inter-site averages: $I_\sigma=\left\langle{c_{i\sigma }^+c_{j\sigma }}\right\rangle$, in addition to the
usual on site averages: $n_\sigma=\left\langle{c_{i\sigma }^+c_{i\sigma }}\right\rangle$, which contribute to the
Stoner field.

For the average kinetic energy of  $+\sigma$ electrons  ($K^\sigma$) one can write
\be
	\left\langle {K^\sigma } \right\rangle =-t_{eff}^\sigma \sum\limits_{<i,j>} 
{\left\langle {c_{i\sigma }^+ c_{j\sigma } } \right\rangle } 
=-zt_{eff}^\sigma I_\sigma=-D_\sigma I_\sigma ,
\label{6}
	\ee  
where $z$ is the number of the nearest neighbors, $D_\sigma=zt_{eff}^\sigma$   is the half band-width of the
$+\sigma$ electrons, and effective hopping integral $t_{eff}^\sigma=t\cdot b^\sigma$ , with $b^\sigma$ defined by
eq. (10) below.

The kinetic energy can be calculated straightforward as 
 \be
	\left\langle {K^\sigma } \right\rangle =\int\limits_{-D }^D 
{f({E^\sigma({\varepsilon}) })E^\sigma({\varepsilon}) \rho _\sigma ({\varepsilon} 
)d\varepsilon}=-D_\sigma I_\sigma  ,
\label{7}	\ee  
with $E^\sigma({\varepsilon})$  given by eq. (9) with omitted index $k$. Eq. (7) means that we can calculate our
average product of two nearest neighbor operators: $I_\sigma\sim \left\langle {K^\sigma } \right\rangle$.		
In the case of zero temperature, $T= \text{0 K}$, and constant DOS: $\rho _\sigma  \left( \varepsilon  \right) =
const = \frac{1}{D}$ for $-D\leq \varepsilon \leq D$, we obtain from eq. (7) that
\be
I_\sigma=n_\sigma \left( 1-n_\sigma )\right.
\label{8}  								
\ee

This approximation will be used later on throughout this paper. 

The modified H-F approximation will lead to the modified dispersion relationship (see \cite{1,7})
 \be
	E_k^\sigma =\varepsilon _k b^\sigma - \sigma \Delta E_\sigma ,
	\label{9}
	\ee  							
with the parameter $b^\sigma$  describing the spin dependent change of the bandwidth given by
	\be
b_\sigma   = 1 - 2\frac{{\Delta t}}{{t}} n_{ - \sigma }  + 2\frac{{t_{ex} }}{{t}}\left( {n_{ - \sigma }^2  - I_{ -
\sigma }^2  - I_\sigma  I_{ - \sigma } } \right) - \frac{{J - V}}{{t }}I_\sigma   - \frac{{J + J'}}{{t }}I_{ - 
\sigma } .
\label{10} 
\ee 
and the effective Stoner shift (coming from all interaction constants standing in front of the single site operator
$\hat {n}_{i\sigma}$) equal to 
\be
\Delta E_\sigma =(F-zJ) \frac{{m}}{{2}}+2zI_{-\sigma}\left( t_{ex} n_ \sigma- \Delta t \right).
\label{11}
\ee

\vskip1.0cm 

\noindent {\Large {\bf 3. The Correlation Effects }} 

\vskip0.5cm

As in the previous paper \cite{8}, we employ the coherent potential approximation to the strong on-site Coulomb
repulsion $U$  and in the result we obtain in Fig. 1 the following schematic transformation of the initial
paramagnetic DOS (see \cite{11}).

\vskip0.5cm 
\begin{center}
Fig. 1
\end{center}
\vskip0.5cm 

After applying the modified H-F approximation to the inter-site interactions we obtain the DOS de-formed by the
inter-site correlation as shown in Fig. 2 (see \cite{2}).  

\vskip0.5cm 
\begin{center}
Fig. 2
\end{center}
\vskip0.5cm

\vskip1.0cm 

\noindent {\Large {\bf 4. The Numerical Results }} 

\vskip0.5cm

The magnetization in Bohr's magnetons is given by 
\be
	m=n_\sigma -n_{-\sigma } ,
	\label{12}
	\ee 								
with the electron occupation numbers for spin $\pm \sigma$ given by the following equation (see \cite{8})   
  \be
 n_{ \pm \sigma }  = \int\limits_{ - \infty }^\infty  {\rho _{ \pm \sigma } \left( \varepsilon 
\right)\frac{{d\varepsilon }}{{1 + e^{\left( {\varepsilon  - \mu  \mp \Delta E_{\pm \sigma} } \right) / {kT} }}}} ,
\label{13}
\ee 
where the spin dependent DOS, $\rho _{ \pm \sigma } \left( \varepsilon  \right)$, is de-formed by both the on-site
and the inter-site correlation, as in Figs 1 and 2.

\vskip0.5cm 
{\bf 4.1 We consider initially the case of on-site Coulomb repulsion $U=0$ }
\vskip0.5cm

{\bf(i) There are only inter-atomic interactions $V$, $J$ and $J'$}

We assume temporarily that the kinetic interactions are zero: $\Delta t=t_{ex}=0$. In this case the total Stoner
field (from eq. (11)) is given by 

   \be
   F_{tot}=F+zJ.
   \label{14}
   \ee								

To limit the number of free parameters we assume also that $J'=J$, and $V=0$. 

The results for the minimum on-site Stoner field $F=F_{tot}-zJ$ are shown in Fig. 3. They show a strong decrease of
the critical on-site Stoner field under the influence of exchange inter-site interaction.

\vskip0.5cm 
\begin{center}
Fig. 3
\end{center}
\vskip0.5cm

Next, we computed from eqs (12) and (13) the temperature at which the value of magnetization drops to zero in the
presence of only the on-site Stoner field $F$ and the inter-site exchange interaction $J$. The results are
collected in Table I, for the values of electron occupation and magnetization corresponding to ferromagnetic 3d
elements. One can see that with the growing component of the on-site field (growing column number from 1 to 4) the
Curie temperature keeps increasing well out of the range of experimental values (column no.5).

\vskip0.5cm 
\begin{center}
TABLE I
\end{center}
\vskip0.5cm

{\bf(ii) the case of nonzero kinetic interactions $\Delta t=t-t_1$ and $t_{ex}=\frac{{t+t_2}}{{2}}-t_1$}

We will assume now that $V=J=J'=0$. In this case the dependence of the on-site critical field $F_{cr}$ on electron
occupation has already been investigated in \cite{7}. The results have shown a strong decrease of the on-site
critical field, with the decreasing hopping 'inhibiting' factor $S$.
This decrease will also cause the decrease of Curie temperatures, as is shown in Table II for different values of
$S$.

\vskip0.5cm 
\begin{center}
TABLE II
\end{center}
\vskip0.5cm

{\bf(iii) both kinetic interactions and inter-atomic interactions are nonzero} 

The kinetic interactions are represented by the hopping 'inhibiting' factor $S$, and the inter-atomic interactions
by $J=J'$  ($V=0$). We still neglect the on-site strong Coulomb repulsion, $U=0$. The results are shown in Fig. 4.
It is evident from this figure that both $S$, and $J$ strongly decrease the on-site Stoner field required to create
ferromagnetism.

\vskip0.5cm 
\begin{center}
Fig. 4
\end{center}
\vskip0.5cm

Therefore we calculated the Curie temperature under the influence of these two factors. The results are shown in
the Table III.

\vskip0.5cm 
\begin{center}
TABLE III
\end{center}
\vskip0.5cm

We can see from Table III that including simultaneously the kinetic interactions ($S$) and the inter-site exchange
interactions ($J$) decreases the Curie temperature close to experimental values (compare columns no.1 and no.5).

\vskip0.5cm 
{\bf4.2 The case of on-site Coulomb repulsion $U\gg D$} 
\vskip0.5cm 

The results in this case without the inter-site interactions are shown in Fig. 5. As can be seen from this figure
there is very little difference in $F_{cr}$ for the realistic $U=3D$, as compared to the case of $U\gg D$.
Therefore, in further calculations, we will use for simplicity large $U\gg D$.

\vskip0.5cm 
\begin{center}
Fig. 5
\end{center}
\vskip0.5cm

We will assume that both kinetic and inter-atomic interactions are nonzero. As previously these inter-site
interactions will be represented by parameters $S$ and $J$. The critical on-site Stoner field calculated in this
case is shown in Fig. 6. The band is being split by the strong Coulomb repulsion, see Fig. 1 for the schematic
depiction of the band. This is why the curves with $S\neq 1$ have a discontinuity at half filling, when the Fermi
energy jumps from the lower to the upper sub-band. Comparing this figure with Fig. 4 shows that the influence of strong Coulomb on-site correlation is not really beneficial for $n\geq 1.4$, where the ferromagnetic 3d elements are located. 

\vskip0.5cm 
\begin{center}
Fig. 6
\end{center}
\vskip0.5cm

To analyze further the influence of $U$ on the magnetic properties we have calculated the magnetization versus
temperature at $U=0$ and $U=\infty$ for the electron occupation representing iron, see Fig. 7.

\vskip0.5cm 
\begin{center}
Fig. 7
\end{center}
\vskip0.5cm

This figure shows that $U$ does not decrease the Curie temperature at concentrations corresponding to existing
ferromagnetic elements. Table IV shows the results of the Curie temperature in the presence of the same interactions as in Table III, with the difference of adding to it a very strong interaction of $U$.

\vskip0.5cm 
\begin{center}
TABLE IV
\end{center}
\vskip0.5cm

Comparing Table III with the Table IV tells us that including the on-site Coulomb correlation in the CPA approximation 
increases slightly the Curie temperature pushing it further away from the experimental values.

\vskip1.0cm 

\noindent {\Large {\bf 5. Conclusions}} 

\vskip0.5cm

First, the direct calculations within the original Stoner model (see \cite{13,14}) have pointed out that the 
Curie temperature, after adjusting the Stoner shift to the experimental magnetic moment at zero temperature, is 
much too high (see Table I, column 6 in here).

As we have found previously \cite{8} the inter-site interactions create the correlation effect (see Fig. 2) already 
in the first order modified H-F approximation. In this paper we introduced the kinetic and the inter-site 
interactions in the first order modified approximation that allowed us to obtain the correct Curie temperatures and 
simultaneously the correct magnetic moments at zero temperature for 3d transition elements (see Table III). 
Considering the simplicity of the model, the results are close enough to the experimental data (see Table III). One 
can write symbolically that 

  \[
\frac{{kT_c }}{{F_{i,j} \left( m \right)}} <  < \frac{{kT_c }}{{F_{i,i} \left( m \right)}},
\]
where $F_{i,j}(m)$ is the inter-atomic Stoner field fitted to the magnetic moment at zero temperature, and 
$F_{i,i}(m)$ is the on-site Stoner field fitted to the magnetic moment at zero temperature. Intuitively speaking we 
can say that the inter-atomic field is much 'softer' and decreases faster with the temperature than the on-site 
Stoner field used in the original Stoner model (see Hubbard \cite{15}).

Hubbard's \cite{15} calculations 'imply that two energy scales are operative in iron, one of the order of electron 
volts which is characteristic of the itinerant behavior (e.g., the bandwidth and the exchange fields), and another 
of the order of one tenth of an electron volt characteristic of the "localized" behavior (e.g., $k_B T_c$, the 
$\Delta E(V)$).' In our model this larger field would be the on-site field creating local moments existing even 
above the Curie temperature, and the smaller field would be the inter-site field responsible for their ordering.

Adding up the on-site strong correlation $U$ in the Section 4.2 (even in the CPA approximation equivalent to the 
sum of some infinite set of diagrams (see \cite{11})), has decreased the critical field, and enhanced magnetic 
susceptibility (see \cite{16}), but did not improve the values of the theoretical Curie temperature (see Table IV). 
This fact can be understood better after examining Fig. 6, where the critical field (initializing magnetization) 
has dropped at half-filling but not at the end of the band, where the 3d elements are located. As was already 
established (e.g. see \cite{17,18}) the electron correlation $U$ can help in creating antiferromagnetism (AF) at 
the half-filled band, where the antiferromagnetic 3d elements (Cr, Mn) are located, by reducing the critical field 
for AF to zero. As was mentioned above, the model is very simple, the details of the realistic DOS could be 
included as well as the magnetization decrease through the spin waves excitation, which would bring the theoretical 
results to complete agreement with the experimental data.

\newpage
{\large {\bf Figure Captions}}

\vskip 0.5cm
\noindent {\bf Fig.1.} Schematic DOS showing the influence of the strong on-site Coulomb correlation, $U$. The 
paramagnetic DOS for both spins, $\pm \sigma$  are solid lines. When $U$ is strong enough, the band is split into 
two sub-bands, lower sub-bands have the capacity of $1-n_{-\sigma}$ for $+\sigma$ electrons, and  $1-n_{\sigma}$ 
for $-\sigma$ electrons. The changes in the spin electron densities integrated over energy (to the Fermi level) are 
shown as the shaded areas in this figure, and they contribute with the sign to the correlation factor $K_U$.

\vskip 0.5cm
\noindent {\bf Fig.2.} Schematic DOS shows the influence of the inter-site interactions. The paramagnetic DOS for 
both spins, $\pm \sigma$ are solid lines. The inter-site interactions change the relative width of the bands with 
respect to each other (described by the $b^\sigma$ factors, see eq. (9)). The Stoner field, which would displace 
the bands with respect to each other, is assumed to be nonzero. The shaded areas in this figure are the inter-site 
correlation factor $K_{ij}$.

\vskip 0.5cm
\noindent {\bf Fig.3.} Dependence of the on-site critical Stoner field, $F_{cr}=F_{tot}^{cr}-zJ$ (in the units of half band-width), on electron occupation, for different exchange inter-site interactions $J$; $J=0$ (Stoner model) - continuous line, $J=0.5t$ - dashed line, $J=0.75t$ - dotted line.

\vskip 0.5cm
\noindent {\bf Fig.4.} Dependence of the on-site critical Stoner field, $F_{cr}$  (in the units of half 
band-width), on the electron occupation, for different values of $S$ and SJ; $S=1$ and $J=0$  (the Stoner model) - dot-dashed line, $S=0.6$  and $J=0.5t$ - solid line, $S=1$ and $J=0.5t$ - dashed line, $S=0.6$ and $J=0$ - dotted line. 

\vskip 0.5cm
\noindent {\bf Fig.5.} Dependence of the on-site critical Stoner field, $F_{cr}$ (in the units of half band-width) on electron occupation, for different on-site Coulomb repulsion $U$ ($J=0$, $S=1$).

\vskip 0.5cm
\noindent {\bf Fig.6.} Dependence of the on-site critical Stoner field, $F_{cr}$ (in the units of half band-width), on electron occupation, shows the influence of inter-site interactions, $J=J'$, $V=0$, and of the hopping interactions, $\Delta t$, $t_{ex}$, represented by the hopping 'inhibiting' factor $S$, in the presence of strong on-site correlation, $U=\infty$.

\vskip 0.5cm
\noindent {\bf Fig.7.} Magnetization (in Bohr's magnetons) dependence on temperature, $m(T)$, with and without the strong Coulomb correlation, $n=1.4$, $D=2.8 \text{eV}$ for different values of parameters $U$, $J$ and $S$; $U=0$, $J=0$, $S=1$- solid line, $U=\infty$, $J=0$, $S=1$ - dashed line, $U=0$, $J=0.5t$, $S=0.6$ - dotted line, $U=\infty$, $J=0.5t$, $S=0.6$ - dot-dashed line.

\newpage
\begin{figure}
\epsfxsize10cm
\epsffile{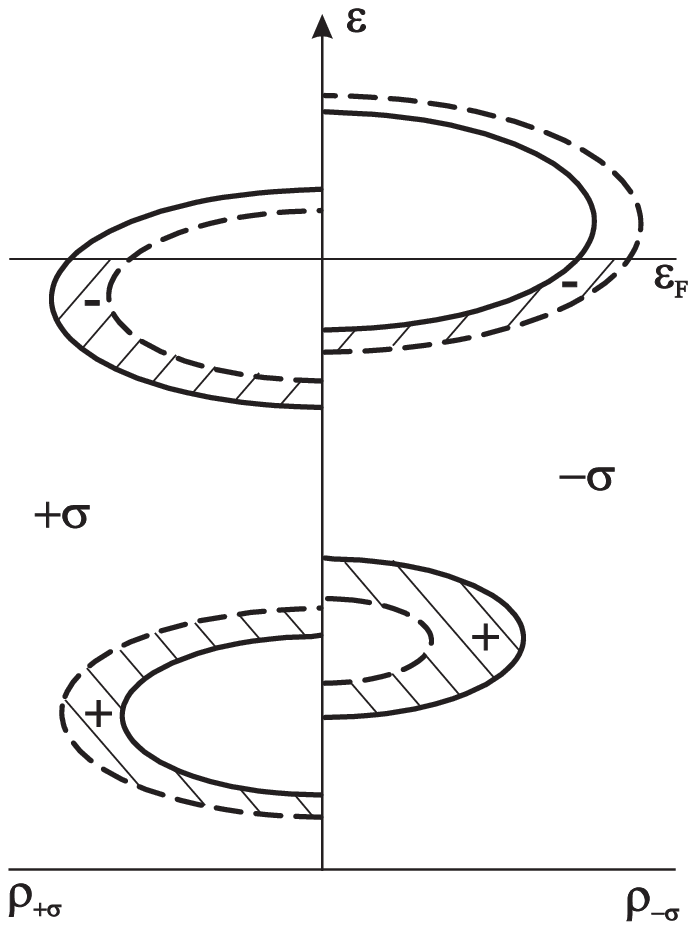}
\caption{}
\end{figure}

\newpage
\begin{figure}
\epsfxsize10cm
\epsffile{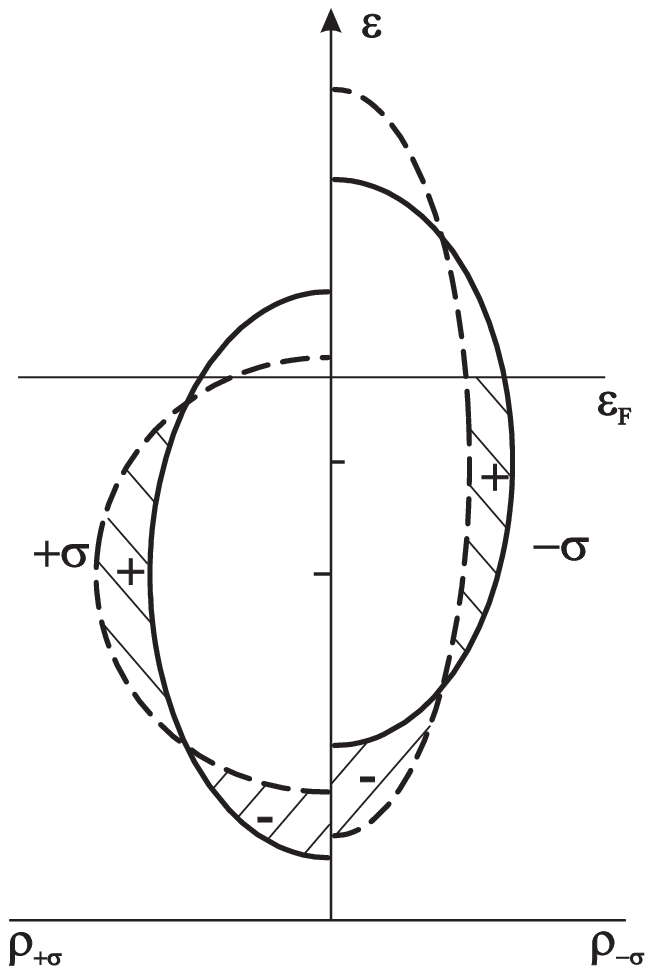}
\caption{}
\end{figure}

\newpage
\begin{figure}
\epsfxsize14cm
\epsffile{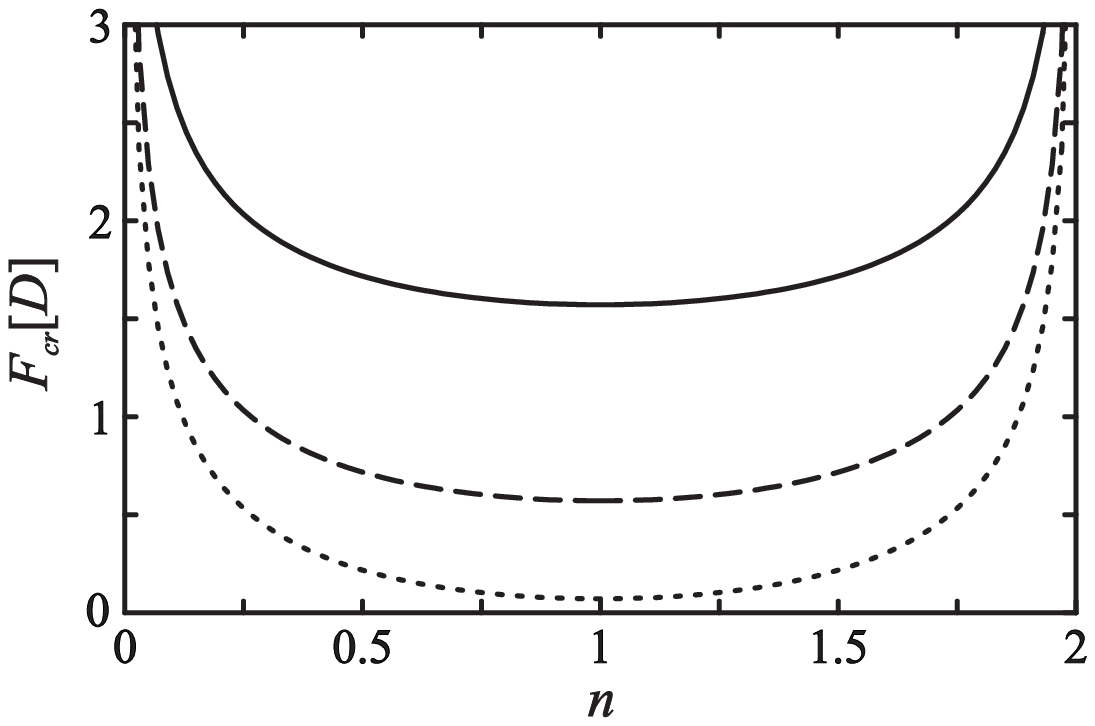}
\caption{}
\end{figure}

\newpage
\begin{figure}
\epsfxsize14cm
\epsffile{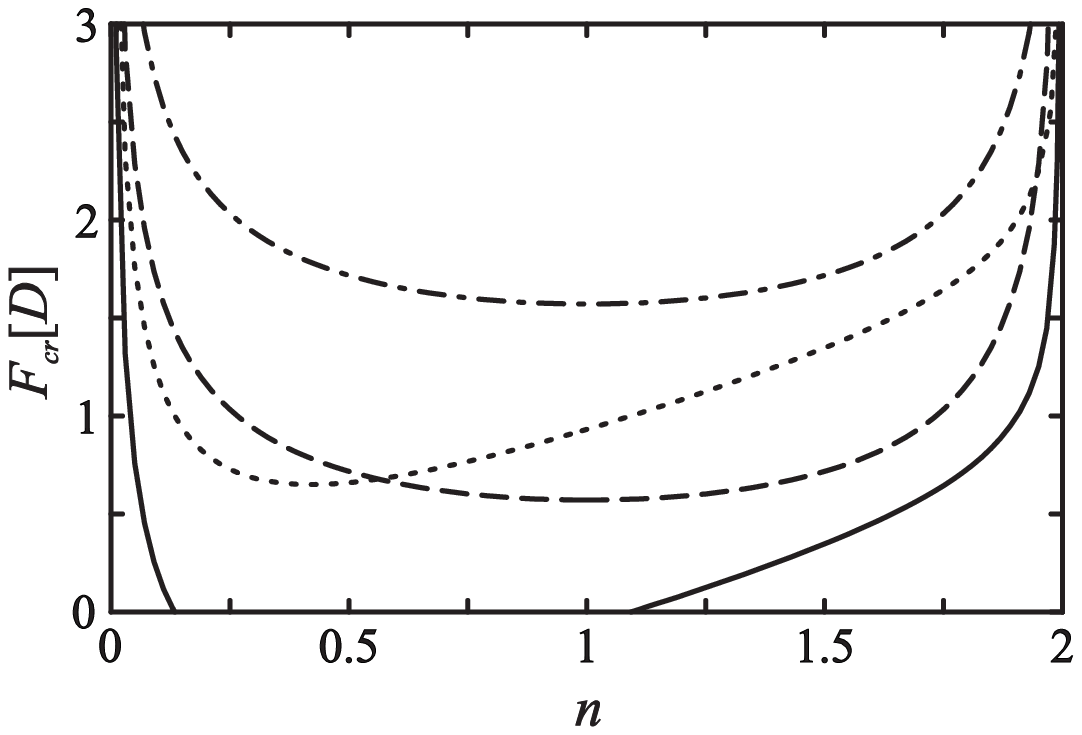}
\caption{}
\end{figure}

\newpage
\begin{figure}
\epsfxsize14cm
\epsffile{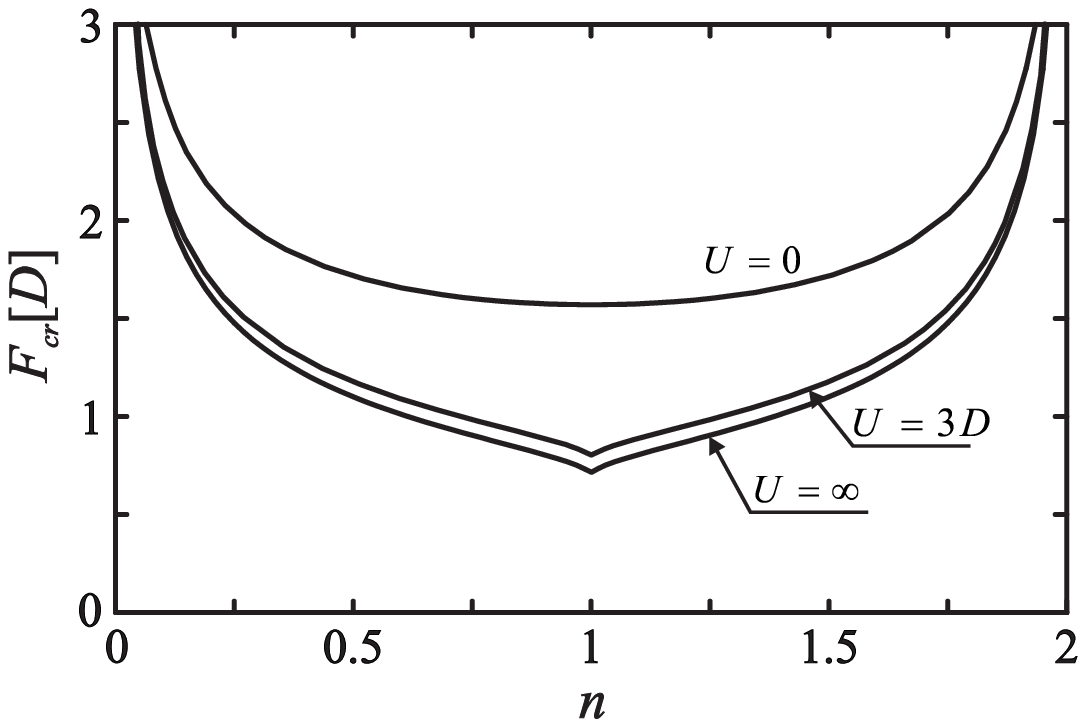}
\caption{}
\end{figure}

\newpage
\begin{figure}
\epsfxsize14cm
\epsffile{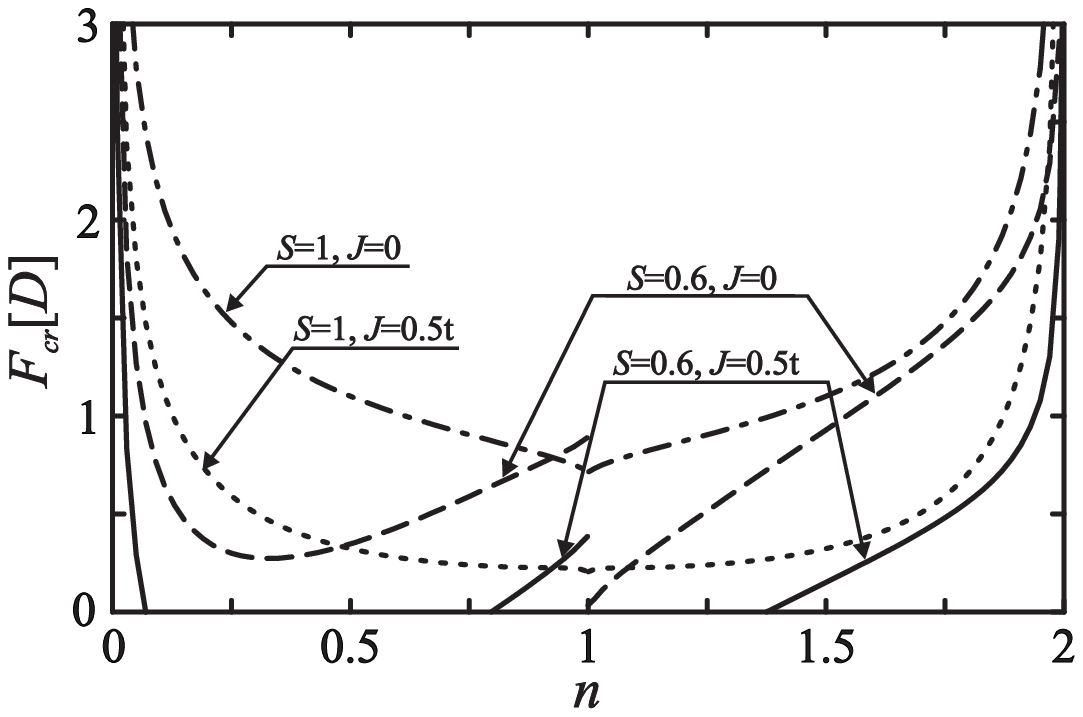}
\caption{}
\end{figure}

\newpage
\begin{figure}
\epsfxsize14cm
\epsffile{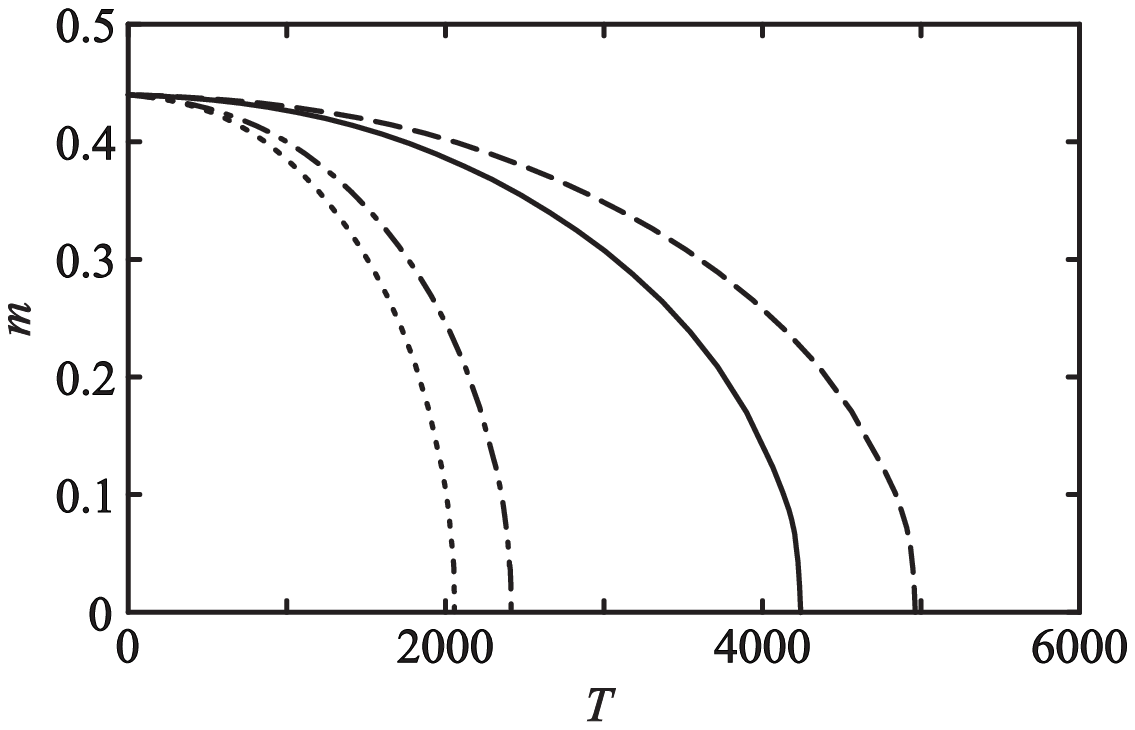}
\caption{}
\end{figure}

\newpage
\begin{table}[b]
\begin{center}
\begin{tabular}{p{14pt}p{28pt}p{28pt}p{34pt}p{56pt}p{59pt}p{70pt}p{56pt}p{32pt}p{32pt}}
\hline
\raisebox{-1.50ex}[0cm][0cm]{El.}& 
\raisebox{-1.50ex}[0cm][0cm]{$n$}& 
\raisebox{-1.50ex}[0cm][0cm]{$m$}&
\raisebox{-1.50ex}[0cm][0cm]{$D$ [eV]}&  
\multicolumn{6}{p{309pt}}{$T_c$ [K]}  \\
\cline{5-10} 
 & 
 & 
 & 
 &
\scriptsize No. 1 \par $F_{tot}=0,$ \par $F=-zJ(m)$& 
\scriptsize  No. 2 \par $F_{tot}=zJ(m)$ \par $F\equiv0$& 
\scriptsize  No. 3 \par $F_{tot}=zJ+F(m)$ \par $J=0.75t$& 
\scriptsize  No. 4 \par $F_{tot}=F(m)$ \par $J\equiv0$&
\scriptsize  No. 5 \par $T_c^{exp }$& 
\scriptsize  No. 6 \par $T_c^{parab}$\\
\hline
Fe& 
1.4& 
0.44& 
2.8&
1750& 
3340& 
3640& 
4290& 
1043& 
4820 \\

Co& 
1.65& 
0.344&
2.65& 
1590& 
3000& 
3500& 
4710& 
1388& 
3380 \\

Ni& 
1.87& 
0.122& 
2.35&
920& 
1380& 
1620& 
1960& 
627& 
1020 \\
\hline
\end{tabular}
\label{tab1}
\end{center}

\caption{ {\bf Curie Temperature for Ferromagnetic Elements, the role of inter-site exchange interaction},
$n$=3d electron number normalized to 2,
$m$=magnetic moment at 0K normalized to 2,
$D$= 3d half band-width according to [12],
columns no.1,2,3,4- $T_c$[K]=Curie temperature for different are for different $F$ and $J$, $F(m)$ and $J(m)$ are fitted to $m$ at 0K,
$T_c^{exp}$ experimental Curie temperature in K,
$T_c^{parab}$=Curie temperature for the Stoner model with parabolic DOS.}

\end{table}

\newpage
\begin{table}[b]
\begin{center}
\begin{tabular}{p{57pt}p{58pt}p{58pt}p{59pt}p{59pt}p{59pt}p{59pt}}
\hline
\raisebox{-1.50ex}[0cm][0cm]{El.}& 
\raisebox{-1.50ex}[0cm][0cm]{$n$}& 
\raisebox{-1.50ex}[0cm][0cm]{$m$}& 
\multicolumn{4}{p{240pt}}{$T_c$ [K]}  \\
\cline{4-7} 
 & 
 & 
 & 
\scriptsize $S=0.2$& 
\scriptsize $S=0.6$ & 
\scriptsize $S=1$ & 
\scriptsize $T_c^{exp }$\\
\hline
Fe& 
1.4& 
0.44& 
2445& 
3295& 
4290& 
1043 \\

Co& 
1.65& 
0.344& 
1580& 
3300& 
4710& 
1388 \\

Ni& 
1.87& 
0.122& 
220& 
870& 
1960& 
627 \\
\hline
\end{tabular}
\label{tab2}
\end{center}

\caption{{\bf Curie Temperatures for Ferromagnetic Elements, the role of kinetic interactions $T_c=T_c(S)$}}

\end{table}

\newpage
\begin{table}[b]
\begin{center}
\begin{tabular}{p{29pt}p{34pt}p{34pt}p{61pt}p{68pt}p{75pt}p{61pt}p{47pt}}
\hline
\raisebox{-1.50ex}[0cm][0cm]{El.}& 
\raisebox{-1.50ex}[0cm][0cm]{$n$}& 
\raisebox{-1.50ex}[0cm][0cm]{$m$}& 
\multicolumn{5}{p{316pt}}{$T_c$ [K]}  \\
\cline{4-8} 
 & 
 & 
 & 
\scriptsize  No. 1 \par $S=0.6$ \par $J=0.5t$& 
\scriptsize  No. 2 \par $S=0.6$ \par $J=0$&  
\scriptsize  No. 3 \par $S=1$ \par $J=0.5t$&  
\scriptsize  No. 4 \par $S=1$ \par $J=0$& 
\scriptsize  No. 5 \par $T_c^{exp}$\\ 
\hline
Fe& 
1.4& 
0.44& 
2050& 
3295& 
3980& 
4290& 
1043 \\

Co& 
1.65& 
0.344& 
1690& 
3300& 
3880& 
4710& 
1388 \\

Ni& 
1.87& 
0.122& 
620& 
870& 
1720& 
1960& 
627 \\
\hline
\end{tabular}
\label{tab3}
\end{center}

\caption{{\bf Curie Temperatures for Ferromagnetic Elements, the role of kinetic and inter-site exchange interactions at $U=0$}, 
no.1,2,3- $T_c(S,J)$= Curie temperature for fixed $J,S$ and $F(m)$, $F(m)$ is fitted to $m$ at 0K,
n0. 4 -$T_c(S=1,J=0)$= Curie temperature for the case of Stoner model, no inter-site interactions $(J=J'=V=\Delta t=t_{ex}\equiv 0)$, $F(m)$ is fitted to $m$ at 0K}
\end{table}

\newpage
\begin{table}[b]
\begin{center}
\begin{tabular}{p{29pt}p{34pt}p{34pt}p{61pt}p{68pt}p{75pt}p{61pt}p{47pt}}
\hline
\raisebox{-1.50ex}[0cm][0cm]{El.}& 
\raisebox{-1.50ex}[0cm][0cm]{$n$}& 
\raisebox{-1.50ex}[0cm][0cm]{$m$}& 
\multicolumn{5}{p{316pt}}{$T_c$ [K]}  \\
\cline{4-8} 
 & 
 & 
 & 
\scriptsize  No. 1 \par $S=0.6$ \par $J=0.5t$& 
\scriptsize  No. 2 \par $S=0.6$ \par $J=0$&  
\scriptsize  No. 3 \par $S=1$ \par $J=0.5t$&  
\scriptsize  No. 4 \par $S=1$ \par $J=0$& 
\scriptsize  No. 5 \par $T_c^{exp}$\\
\hline
Fe& 
1.4& 
0.44& 
2420& 
4000& 
3810& 
4960& 
1043 \\

Co& 
1.65& 
0.344& 
1850& 
3270& 
3990& 
5180& 
1388 \\

Ni& 
1.87& 
0.122& 
630& 
910& 
1770& 
2020& 
627 \\
\hline
\end{tabular}
\label{tab4}
\end{center}

\caption{{\bf Curie Temperatures for Ferromagnetic Elements, the role of kinetic and inter-site exchange interactions at $U=\infty$}}
\end{table}

\end{document}